\documentclass{article}
\usepackage{spconf,amsmath,graphicx,hyperref}
\usepackage{enumitem,booktabs,subcaption, soul, xcolor}
\soulregister{\cite}{7}
\usepackage{caption, amsmath, amssymb}
\usepackage{cite}
\title{UnwrapDiff: A Conditional Diffusion Model for InSAR Phase Unwrapping}
\name{%
  Yijia~Song$^{1,3}$ 
  Juliet~Biggs$^{2}$ 
  Alin~Achim$^{1}$ 
  Robert~Popescu$^{1,3}$ 
  Simon~Orrego$^{2}$ 
  Nantheera~Anantrasirichai$^{1,3}$%
}
\address{%
  $^{1}$ Visual Information Laboratory, University of Bristol, Bristol, UK\\
  $^{2}$ COMET, School of Earth Sciences, University of Bristol, Bristol, UK\\
  $^{3}$ COMET, School of Computer Science, University of Bristol, Bristol, UK\\
}

\begin{document}
%
\maketitle
\begin{abstract}
Phase unwrapping is a fundamental problem in InSAR data processing, supporting geophysical applications such as deformation monitoring and hazard assessment. Its reliability is limited by noise and decorrelation in radar acquisitions, which makes accurate reconstruction of the deformation signal challenging. We propose a denoising diffusion probabilistic model (DDPM)-based framework for InSAR phase unwrapping, UnwrapDiff, in which the output of the traditional minimum cost flow algorithm (SNAPHU) is incorporated as conditional guidance. To evaluate robustness, we construct a synthetic dataset that incorporates atmospheric effects and diverse noise patterns, representative of realistic InSAR observations. Experiments show that the proposed model leverages the conditional prior while reducing the effect of diverse noise patterns, achieving on average a 10.11\% reduction in NRMSE compared to SNAPHU. It also achieves better reconstruction quality in difficult cases such as dyke intrusions. 
\end{abstract}
\section{Introduction}
\label{sec:intro}

Interferometric Synthetic Aperture Radar (InSAR) is a powerful technique for monitoring ground deformation related to earthquakes, volcanic activity and other geophysical processes \cite{Poland2022Volcano, Ebmeier2018Synthesis, popescu2025unsupervised}. A key step in InSAR processing is phase unwrapping. Radar phase measurements are recorded modulo 2$\pi$, which introduces discontinuities in the observed signal. The unwrapped phase represents the true ground deformation and can be directly used for subsequent geophysical analysis. The goal is to reconstruct the absolute interferometric phase from wrapped observations. The problem is inherently ill-posed, because both the continuous phase and the integer ambiguity must be determined. Accurate unwrapping is essential for reliable deformation mapping. However, it remains challenging in certain cases, particularly where noise, decorrelation, or sharp deformation gradients occur~\cite{Yan2025A}.

\begin{figure}[t]
    \centering    
    \includegraphics[width=\columnwidth]{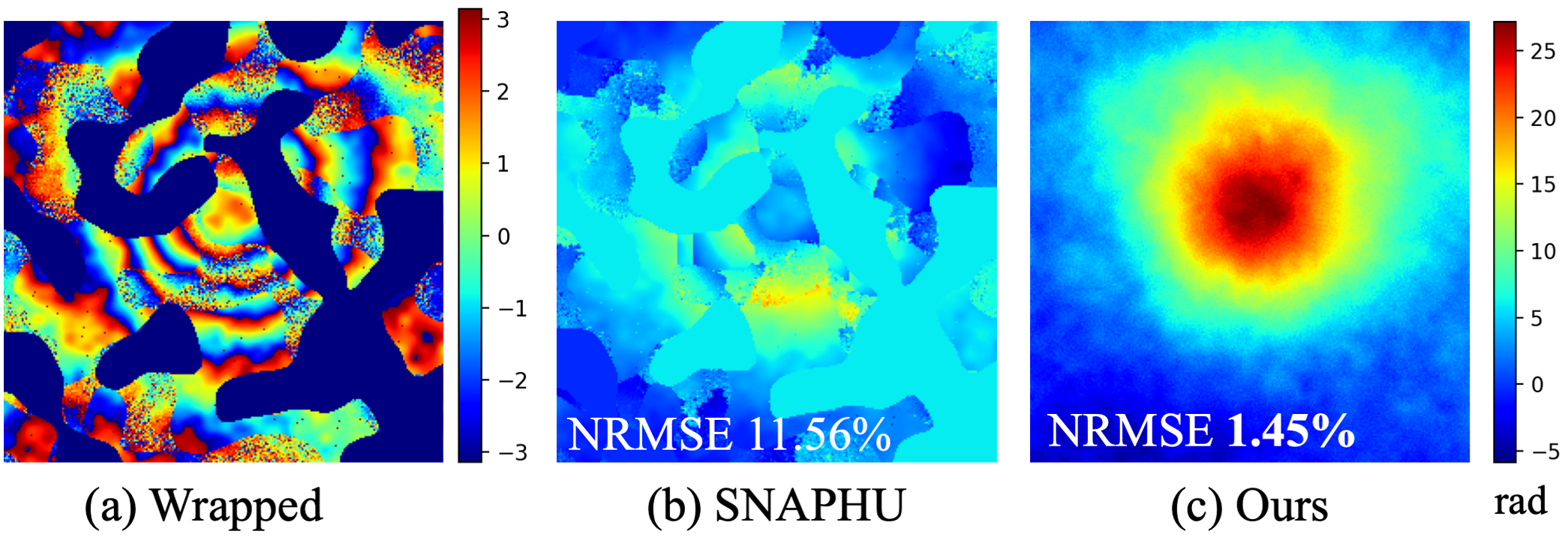} 
    \caption{Unwrapping results of a dyke intrusion scene with water body patches, shown as dark blue regions in the wrapped interferogram.}
    \label{fig:showPU}
\end{figure}

Traditional phase unwrapping methods can be grouped into path-following and optimization-based approaches. Path-following methods, such as branch-cut~\cite{goldstein1998radar}, unwrap phase by integrating along paths while cutting around residues. Optimization-based methods, such as Statistical-cost, Network-flow Algorithm for Phase Unwrapping (\textsc{SNAPHU})~\cite{chen2002phase}, solve a minimum-cost flow problem to enforce global consistency. These approaches are effective in simple cases, but they perform poorly in low-coherence regions (e.g. water bodies) or under sharp deformation gradients (as shown in Fig.~\ref{fig:showPU}), where discontinuities and incorrect branch cuts often appear. 

Deep learning has only recently been applied to InSAR phase unwrapping, and the number of task-specific designs remains limited. PhaseNet~\cite{phasenet} first formulated unwrapping as a multi-class classification task by predicting wrap counts, and was later extended by PhaseNet 2.0~\cite{phasenet2}, which improved the network architecture and training objectives to enhance robustness. U3Net~\cite{U3NET} introduces an unsupervised strategy using deep unrolling that leverages coherence-driven and reconstruction losses to reduce dependence on paired ground truth. SQD-LSTM~\cite{SQD} captures sequential dependencies in the wrapped-to-unwrapped mapping through recurrent modeling. Beyond methods designed specifically for unwrapping, the networks originally designed for image restoration have also been adapted to this task. 
Transformer-based architectures, such as Restormer~\cite{Restormer}, have been applied to phase unwrapping by modifying their loss functions~\cite{U3NET}. More recently, generative approaches have been considered~\cite{Zhou2022PU-GAN}. Diffusion probabilistic models~\cite{ho2020ddpm,nichol2021improved} learn to reverse a progressive noising process and have achieved state-of-the-art performance in image restoration. Although diffusion models have shown high potential in image generation and reconstruction tasks, their potential for phase unwrapping remains unexplored. 

In this paper, we address the aforementioned challenges by proposing \textbf{UnwrapDiff}, a conditional diffusion framework for InSAR phase unwrapping. The method uses SNAPHU outputs as priors to provide global consistency while correcting local errors in noisy and high-gradient regions as shown in Fig.~\ref{fig:showPU}c. Diffusion models offer unique advantages for this task: their iterative denoising process not only captures local textures and global spatial structures, but also enables stable reconstruction across phase discontinuities. This property is consistent with their demonstrated success in image inpainting and missing-data restoration, where diffusion-based methods recover continuous spatial fields despite gaps, noise, or severe corruption. Compared to direct regression or discriminative networks, diffusion models are therefore particularly suited to enforcing spatial constraints and global consistency in phase unwrapping. 

To enable systematic evaluation, we construct a synthetic dataset that integrates deformation, atmospheric effects, and multiple noise patterns to enable systematic robustness evaluation. Experiments demonstrate that the proposed approach achieves higher reconstruction accuracy than exiting methods, particularly under decorrelation and complex deformation. These results suggest that diffusion models have strong potential as a robust alternative for InSAR phase unwrapping.

\section{Methodology}
\label{sec:Methodology}

\subsection{Preliminaries}
\noindent \textbf{Phase Unwrapping Problem}:
Phase unwrapping seeks to recover a continuous phase field from wrapped interferometric observations. 
For an interferogram $\phi_{\text{wrap}} \in (-\pi, \pi]$, the objective is to reconstruct the true unwrapped signal $\phi$. 
Formally, the wrapped measurement is expressed as
\begin{equation}
\phi_w(x) = \mathrm{mod}(\phi(x) + \pi, 2\pi) - \pi,
\end{equation}
where $\phi(x)$ is the continuous phase and $\phi_w(x)$ its wrapped counterpart. In practice, $\phi_w(x)$ is the observed signal, while $\phi(x)$ is the unknown quantity to be recovered. Algorithms such as SNAPHU address this problem by estimating the correct integer multiple of 2$\pi$ across all pixels through global optimization strategies.

This can be equivalently viewed as recovering an integer offset field that links the wrapped and unwrapped phases. In the ideal case, the phase gradient is preserved under wrapping, i.e.,
$\nabla \phi(x) \approx \nabla \phi_w(x)$.
In practice, the phase gradient of the wrapped and unwrapped signals should ideally be consistent. Discontinuities such as wrapping between $-\pi$ and $\pi$, noise, and missing values often violate this assumption. As a result, inconsistencies arise in the gradient field. These inconsistencies appear as residues, defined as non-zero circulation of the wrapped gradient around a closed loop~\cite{Gao2025A}. Addressing residues in a globally consistent manner renders phase unwrapping an ill-posed and challenging problem.

\vspace{2mm}
\noindent \textbf{SNAPHU Algorithm}:
Among classical approaches, the SNAPHU~\cite{chen2002phase} is one of the most widely used.
It partitions an interferogram into smaller tiles and unwrapped individually. Within each tile, irregularly shaped, independent regions of high reliability are identified, and the relative phase offsets between these regions are resolved through a secondary optimization problem. This problem maximizes the a posteriori probability of the full-phase estimate and is formulated as a network flow, minimizing a cost function:
\begin{equation}
\min_{f_{ij} \in \mathbb{Z}} \sum_{(i,j)} c_{ij} f_{ij},
\end{equation}
where $f_{ij}$ denotes the integer flow along edge $(i,j)$ and $c_{ij}$ is its associated cost. 
The cost is derived from statistical models of noise and coherence, so edges in low-coherence regions are penalized less. 
By solving this optimization, \textsc{SNAPHU} enforces global consistency. However, its performance degrades in practice when phase gradients are large or when strong noise is present, often leading to unwrapping errors.

\subsection{Conditional Diffusion Framework}

\begin{figure}[t]
    \centering    \includegraphics[width=\columnwidth]{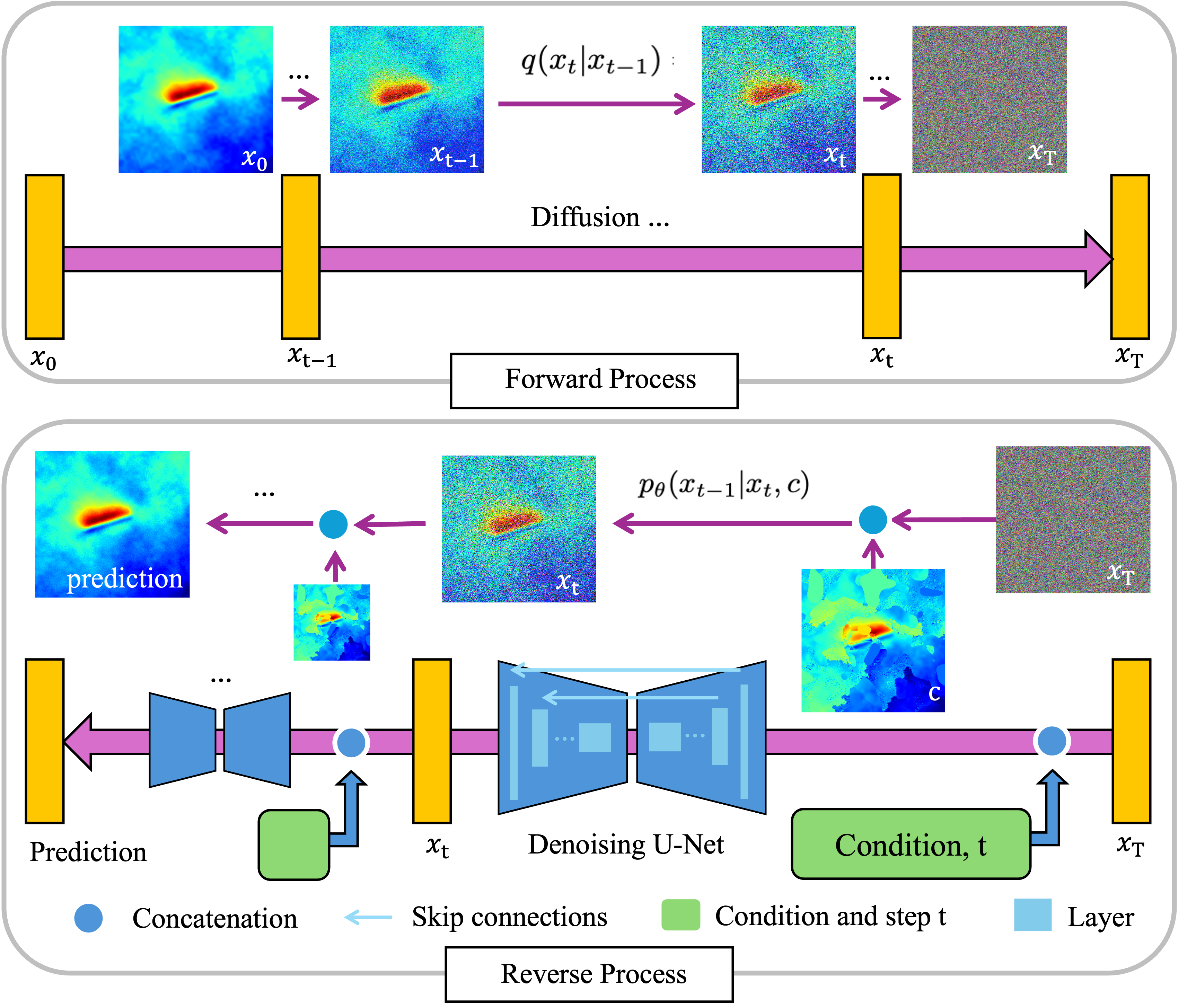} 
    \caption{Diagram of the proposed UnwrapDiff.}
    \label{fig:CDF}
\end{figure}

The proposed method adopts an improved denoising diffusion probabilistic model (DDPM) for InSAR phase unwrapping. 
The design incorporates the output of \textsc{SNAPHU} as a conditional prior, as shown in Fig.~\ref{fig:CDF}. 
This prior provides global consistency while allowing the diffusion model to correct local errors in noisy or high-gradient regions.

The forward process gradually perturbs the clean unwrapped phase $\phi$ ($x_0=\phi$) into a noisy variable $x_T$ via a Markov chain:
\begin{equation}
q(x_t | x_{t-1}) = \mathcal{N}(x_t; \sqrt{1-\beta_t}\, x_{t-1}, \beta_t \mathbf{I}),
\end{equation}
with a cosine variance schedule $\{\beta_t\}_{t=1}^T$.
After $T$ steps the distribution approaches an isotropic Gaussian.

The reverse process reconstructs the clean signal from Gaussian noise, while the SNAPHU prior $c$ is incorporated into the U-Net as an additional input channel, enabling the model to leverage this prior during denoising:
\begin{equation}
p_\theta(x_{t-1} | x_t, c) = \mathcal{N}\!\big(x_{t-1}; \mu_\theta(x_t, t, c), \sigma^2_\theta(x_t, t)\mathbf{I}\big).
\end{equation}
Here $\mu_\theta$ and $\sigma^2_\theta$ are predicted by the network, where $\sigma^2_\theta$ is parameterized between fixed bounds and trained jointly. 
This formulation enables the model to leverage the prior from \textsc{SNAPHU} while maintaining flexibility to correct its systematic errors.

\subsection{Training and Loss Functions}

Training requires paired wrapped and unwrapped samples. 
The synthetic dataset integrates deformation models, atmospheric turbulence, stratified delays, and multiple noise types, 
providing a diverse and controlled environment for robustness evaluation.  

Following the improved DDPM formulation~\cite{nichol2021improved}, 
the model is trained to predict the noise $\epsilon$ added at each step of the forward process. 
The training objective is a weighted mean squared error:
\begin{equation}
\mathcal{L}_{\text{DDPM}} = 
\mathbb{E}_{t, x_0, \epsilon}\left[w_t \cdot 
\left\| \epsilon - \epsilon_\theta(x_t, c, t) \right\|_2^2 \right],
\end{equation}
where $x_t$ is the noisy sample at step $t$, $c$ denotes the \textsc{SNAPHU} prior, 
and $w_t$ is a variance-schedule-dependent weight.  

This objective directly follows the improved diffusion framework and ensures stable and efficient training, 
while the conditional prior $c$ provides additional guidance to correct systematic errors from \textsc{SNAPHU}. 

\section{Experimental results}
\label{sec:Experiment}

\subsection{Dataset}
The synthetic dataset is generated by combining three key components~\cite{anantrasirichai2019deep}: surface deformation, stratified atmospheric delays, and turbulent atmospheric artefacts. Deformation signals are derived from elastic source models, stratified delays from GACOS products, and turbulence from statistical simulations based on Sentinel-1 data. Interferograms are formed as linear combinations of these components and wrapped into $(-\pi, \pi]$, providing a controlled yet realistic benchmark for evaluation~\cite{ansari2019insar}. In addition, three types of noise frequently encountered in InSAR observations are incorporated, namely pepper, speckle, and patchy noise. Pepper noise, owing to its straightforward nature, is not further elaborated. The dataset comprises 11,000 samples in total, with 10,000 used for training and 1,000 reserved for testing.

\textbf{Speckle noise generation.} 
We follow the Goodman statistical model to simulate correlated single-look complex (SLC) pairs\cite{Goodman1984}. 
Given a clean phase field $\phi(\mathbf{x})$ and amplitude $A(\mathbf{x})$, two correlated samples $(z_1, z_2)$ are generated with local coherence $\gamma(\mathbf{x})$. 
The interferogram is then formed as $I = z_1 z_2^\ast$, and the observed phase $\hat{\phi} = \arg(\bar{I})$ after $L$-look averaging represents the speckle-contaminated signal. 
This ensures physically consistent second-order statistics, with phase variance approximately 
$\operatorname{Var}(\hat{\phi}) \approx \tfrac{1-|\gamma|^2}{2L|\gamma|^2}$ under high-coherence conditions\cite{Sintes2012Empirical}.

\textbf{Patchy noise generation.} 
Patchy distortions are simulated by thresholding a smoothed Gaussian random field to create irregular low-coherence regions. 
Within each patch, phase values are either replaced with random jumps of $2k\pi$, mimicking decorrelation effects near water bodies or vegetation boundaries. 
This produces spatially clustered corruption patterns, distinct from pixel-wise noise such as speckle.  

\begin{figure*}[t]
    \centering
    \includegraphics[width=\textwidth]{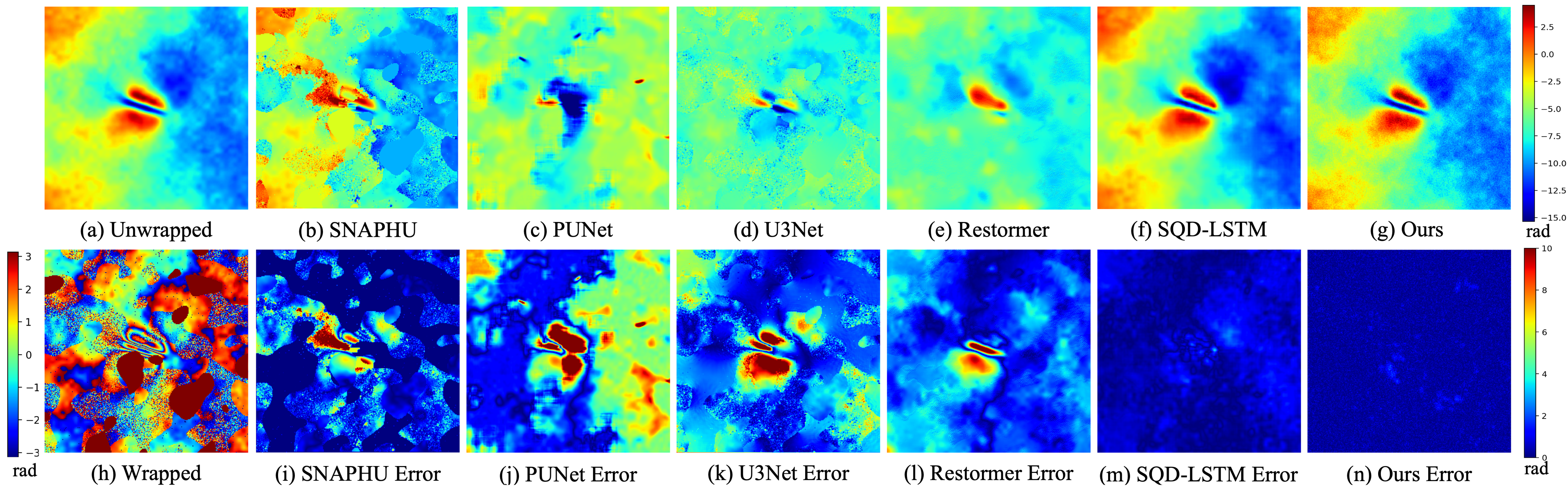} 
    \caption{Unwrapping results (top row) on a simulated dyke intrusion, with the ground truth shown in (a) and the input wrapped interferogram shown in (h). Corresponding error maps are shown in the bottom row.}
    \label{fig:dykeresult}
\end{figure*}

\subsection{Comparative Evaluation}
We compared the proposed method against a range of baselines, including both traditional and learning-based approaches. We compare against both optimization-based method (SNAPHU~\cite{chen2002phase}) and learning-based methods, including PUNet~\cite{punet}, U3Net~\cite{U3NET}, SQD-LSTM~\cite{SQD}, and Restormer~\cite{Restormer}. 
Except for PUNet, which is evaluated with official pretrained weights, all models are retrained on our synthetic dataset for fair comparison.The performance is reported in normalized root mean square error (NRMSE). Lower values indicate better performance.

The results in Table~\ref{tab:comparison} and Fig.\ref{fig:dykeresult} show that SNAPHU performs well in clean regions, but it lacks mechanisms to handle noise. PUNet generates globally smoothed solutions, which reduce atmospheric artefacts but fail to reconstruct sharp dyking deformation. U3Net provides only partial recovery, and its unsupervised design struggles with strong gradients. Restormer restores some deformation in low-noise areas, but it breaks down under severe atmospheric effects. SQD-LSTM achieves good overall unwrapping, yet error maps reveal local artefacts. In contrast, our  UnwrapDiff achieves the lowest error, as it leverages SNAPHU priors while correcting local inconsistencies, leading to accurate recovery of both global patterns and fine details. 

\begin{figure}  
    \centering
    \includegraphics[width=\columnwidth]{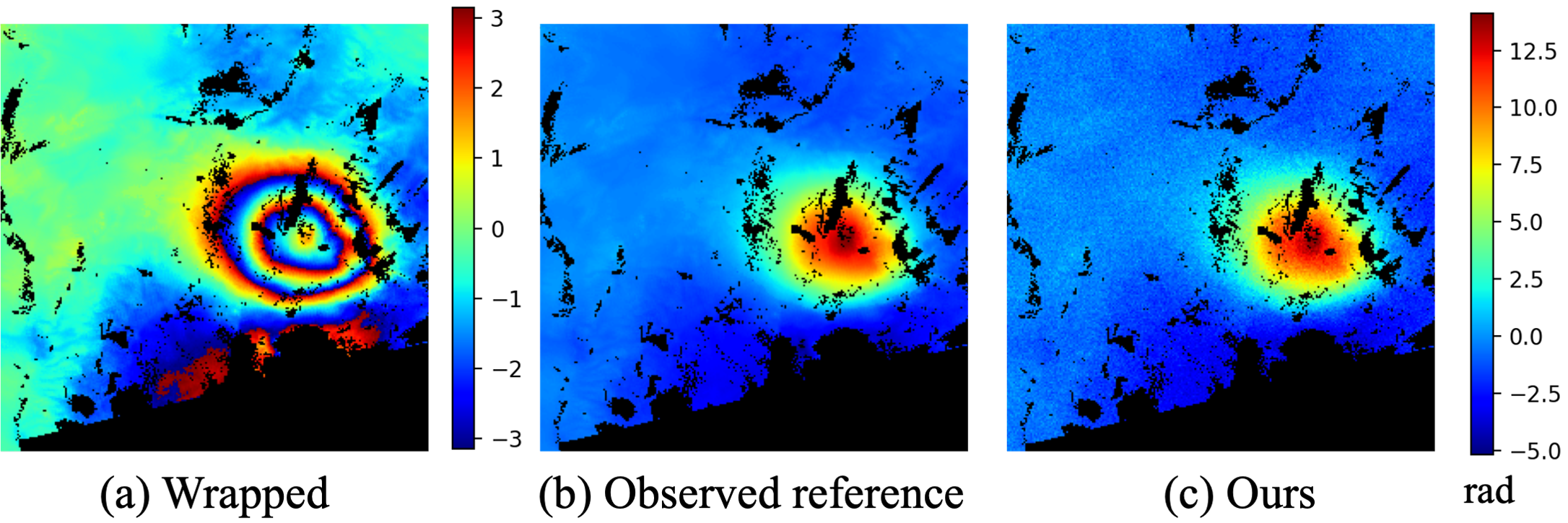} 
    \caption{Unwrapped results from real wrapped interferogram of the 2020 Humahuaca earthquake, Argentina. }
    \label{fig:realinterf}
\end{figure}

\begin{table}[t]
\centering
\caption{Comparison of different phase unwrapping methods based on NRMSE. 
Lower values indicate better performance.}
\label{tab:comparison}
\small
\begin{tabular}{l c}
\toprule
Method & NRMSE (\%)$\downarrow$ \\
\midrule
SNAPHU~\cite{chen2002phase}       &  11.56 \\
PUNet~\cite{punet}                &  14.30 \\
U3Net~\cite{U3NET}                &  18.47 \\
Restormer~\cite{Restormer}        &  10.69 \\
SQD-LSTM~\cite{SQD}               &  3.39 \\
Ours       &  \textbf{1.45} \\
\bottomrule
\end{tabular}
\end{table}
We also conducted validation on real InSAR observations, using a Sentinel-1 TOPS-SAR ascending interferogram capturing the coseismic deformation of the 29 November 2020 Mw 5.7 Humahuaca earthquake in the central Andes of Argentina. The wrapped and unwrapped phases were obtained from previously published results~\cite{orrego2020humahuaca}, where the coseismic displacement field was reconstructed from InSAR time-series spanning two years before and after the earthquake. Due to the lack of GNSS observations for validation, this time-series reconstruction provides a pseudo-ground truth against which unwrapping performance can be evaluated. On this event, our model achieved an NRMSE of 1.97\% showed in Fig\ref{fig:realinterf}, demonstrating its capability to generalize from synthetic training data to real seismic deformation scenarios.

\subsection{Robustness Analysis}
We evaluate the robustness of UnwrapDiff by varying noise levels. 
Table \ref{tab:robustness} demonstrates that SNAPHU exhibits a clear sensitivity to increasing noise levels, with reconstruction errors growing rapidly as corruption intensifies. In contrast, our UnwrapDiff shows only marginal variation across noise severities in both speckle and patchy noises. 
These results confirm that the conditional diffusion design generalizes well to noise and preserves fidelity where optimization-based approaches struggle.

\subsection{Ablations Study}
\textbf{i) Training with only wrapped interferograms:} The model yields an NRMSE of 2.92\%, indicating limited reconstruction accuracy due to the inherent difficulty of resolving phase ambiguities from raw observations alone.
\textbf{ii) Introducing SNAPHU outputs as conditioning priors:} This significantly reduces the error to 1.45\%, demonstrating that classical optimization results provide valuable global constraints, which can be effectively refined by the diffusion process. This highlights the critical role of conditional guidance in enhancing robustness and accuracy in phase unwrapping.

\begin{table}[t]
\centering
\caption{Robustness analysis under varying noise proportions 
for speckle and patchy corruption.}
\label{tab:robustness}
\small
\begin{tabular}{lccc|ccc}
\toprule
& \multicolumn{3}{c}{Speckle noise} & \multicolumn{3}{c}{Patchy noise} \\
\cmidrule(lr){2-4} \cmidrule(lr){5-7}
Method & 35\% & 45\% & 65\% & 10\% & 20\% & 40\% \\
\midrule
SNAPHU    & 6.57 & 9.34 & 14.67 & 2.74 & 4.19 & 8.50 \\
Ours  & 1.25 & 1.27 & 1.34 & 1.28 & 1.34 & 1.66 \\
\bottomrule
\end{tabular}
\end{table}

\section{conclusions}

This paper presents a conditional diffusion framework for InSAR phase unwrapping, which integrates traditional optimization outputs as conditioning priors and refines them through a data-driven diffusion process. Experimental evaluations demonstrate that the framework substantially outperforms both classical and learning-based baselines under diverse noise conditions, achieving superior accuracy while preserving both global consistency and local detail. Our conditional priors enhance precision and robustness across varying noise types and intensities. 

\bibliographystyle{IEEEbib}
\bibliography{refs}

\end{document}